\documentclass[epj,twocolumn]{webofc}
\usepackage[varg]{txfonts}   
\usepackage{graphicx}
\usepackage{float}
\wocname{epj}
\woctitle{Seismology of the Sun and the Distant Stars 2016}
\begin{document}
\title{Isochrones of M67 with an Expanded Set of Parameters}
%

\author{\firstname{Lucas} \lastname{Viani}\inst{1}\fnsep\thanks{\email{lucas.viani@yale.edu}} \and
        \firstname{Sarbani} \lastname{Basu}\inst{1}\ 
}

\institute{Department of Astronomy, Yale University, P.O. Box 208101, New Haven, CT 06520-8101, USA}

\abstract{%
We create isochrones of M67 using the Yale Rotating Stellar Evolution Code. In addition to metallicity, parameters that are traditionally held fixed, such as the mixing length parameter and initial helium abundance, also vary. The amount of convective overshoot is also changed in different sets of isochrones. Models are constructed both with and without diffusion. From the resulting isochrones that fit the cluster, the age range is between 3.6 and 4.8 Gyr and the distance is between 755 and 868 pc. We also confirm Michaud et al. (2004) claim that M67 can be fit without overshoot if diffusion is included.
}
\maketitle
%
\section{Introduction}
\label{intro}
A better understanding of stellar properties and ages is crucial to many fields of astronomy. One widely used method of studying stellar clusters is isochrone fitting. Since stars in a stellar cluster formed around the same time and under the same physical conditions, they have a very small age and metallicity distribution. As a result, the cluster properties (for example age, distance, reddening, and metallicity) can be estimated through the fitting of isochrones. The properties of the isochrone that best fit the cluster's CMD can be used to estimate the properties of the cluster itself. While stellar ages and other properties determined through isochrone fitting are not as precise as asteroseismic methods, obtaining a basic age and property estimate (i.e. mass) of stars helps us later know what frequencies to expect. Here we will generate and fit isochrones to the M67 cluster data, allowing for a wider range of free parameters in our isochrone construction than is typically used.

The M67 cluster (NGC 2682) is the nearest open cluster that is around the solar age. It is estimated to be between 800 and 900 pc away (e.g. as in Sarajendini et al., 2009) at an age of around 4 Gyr (Richer et al., 1998). The cluster has a metallicity around solar, with $\mathrm{[Fe/H]}=0.00\pm0.06$ (Heiter et al., 2014) and a reddening of $E(B-V)=0.041\pm0.004$ (Taylor 2007). A color magnitude diagram (CMD) of M67 can be seen in Figure 1 (data from Geller et al., 2015). It is of importance to note that the M67 CMD has a distinctive ``hook'' feature below the main sequence turn-off. Any isochrone produced must also have this feature in order to be considered a reasonable fit to the cluster.
\begin{figure}[h]
\centering
\includegraphics[width=\hsize,clip]{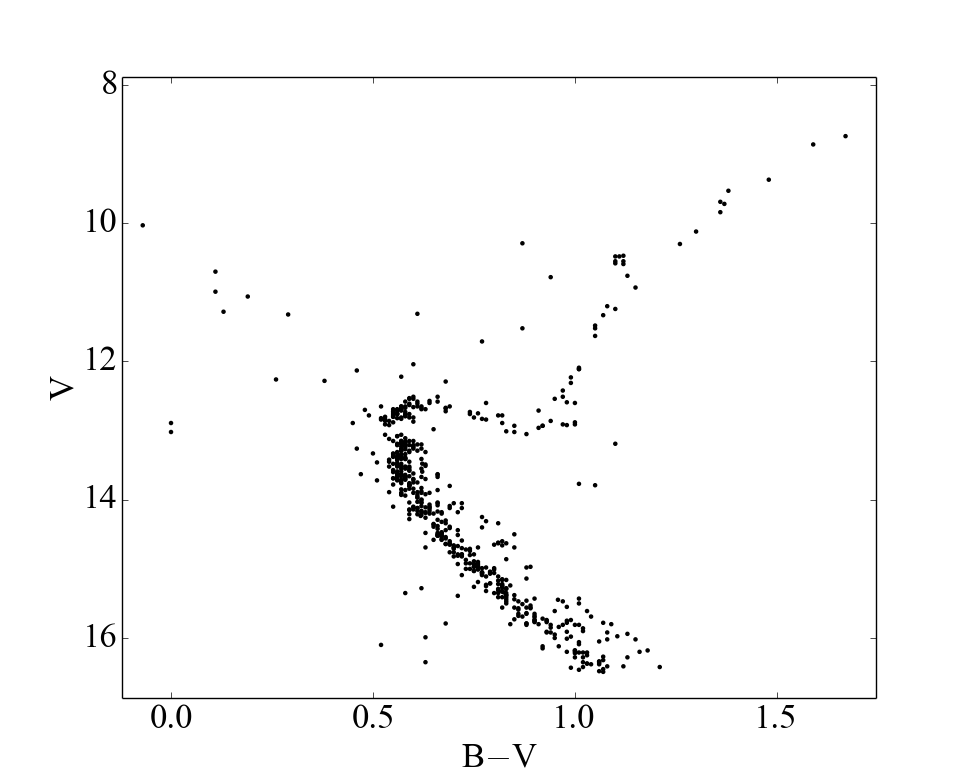}
\caption{Color magnitude diagram for M67, data from Geller et al. (2015). Note the hook-like feature below the main sequence turn-off.}
\label{fig-1}       
\end{figure}


\begin{figure*}[ht]
\centering
\includegraphics[width=\hsize,clip]{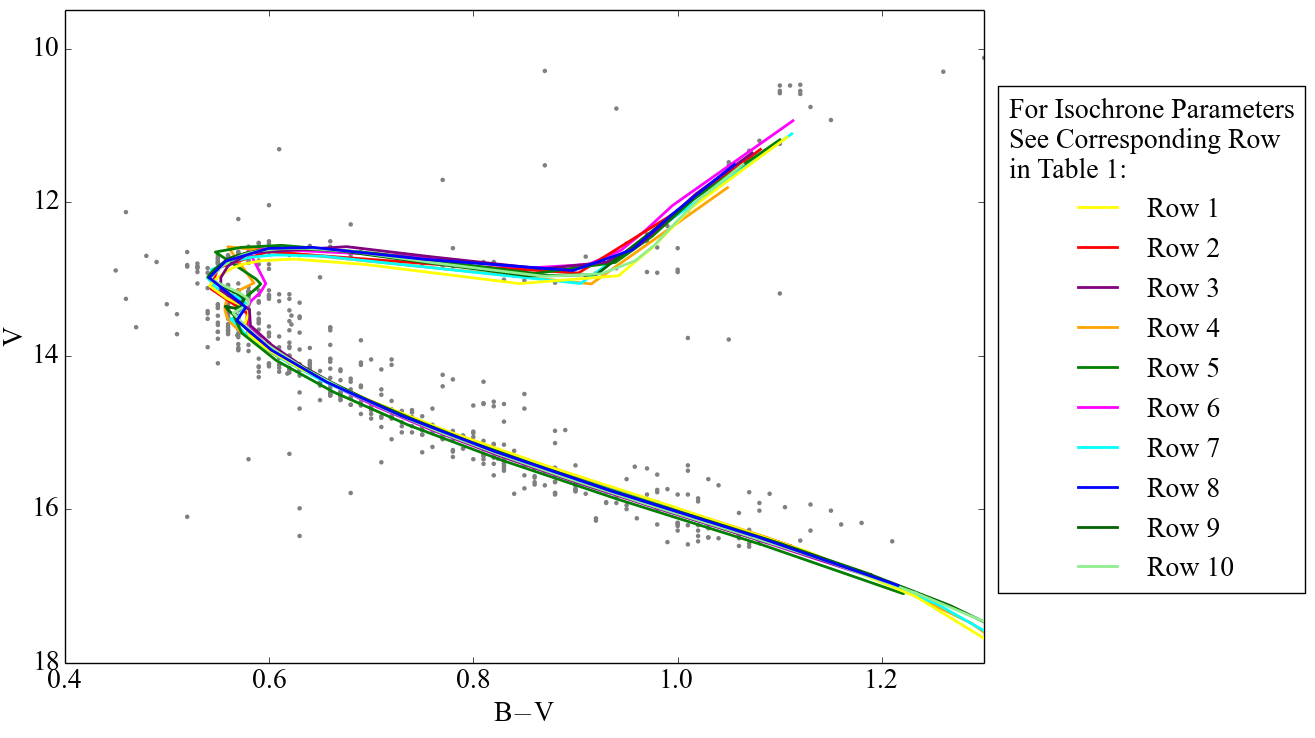}
\caption{The isochrones that fit the M67 CMD. To be a reasonable fit isochrones had to fit the main sequence, turn-off, giant branch, and reproduce the hook-like feature in the CMD. There were 10 such isochrones that fit the M67 CMD. The legend refers to rows in Table 1 where the parameters of the isochrones are listed.}
\label{fig-2}       
\end{figure*}


While many resources currently exist for the quick generation of a desired set of isochrones, such as the YY isochrones (Yi et al., 2001), PARSEC isochones (Bressan et al., 2012), and BaSTI isochrones (Pietrinferni et al., 2004), these methods are limited in the number of parameters that the user can fine-tune and control. For example, the mixing length, initial helium abundance, and the inclusion and amount of diffusion or convective overshoot. As a result, we generated our own set of isochrones through the creation of stellar models, allowing these parameters which are typically held fixed to vary.

The difficulties in accurately modeling convection led to the use of the ``mixing length theory" approximation (B{\"o}hm-Vitense 1958). Here the mean-free path that a convective element travels is determined by $\alpha H_{\mathrm{P}}$ where $H_{\mathrm{P}}$ is the pressure scale height and $\alpha$ is the mixing length parameter. Traditionally, the value of $\alpha$ is determined by creating a solar model. To do this, a relation between metallicity, $Z$, and $Y_0$ is assumed. Then the value of $\alpha$ is adjusted until the model has a luminosity of 1 L$_\odot$ and radius 1 R$_\odot$ at the solar age. Typically isochrones are then created with the values of $\alpha$ and $Y_0$ held fixed. This investigation however will allow both $\alpha$ and $Y_0$ to be free parameters.

It is also of interest to examine the need for convective overshoot or diffusion in M67. Michaud et al. (2004) shows that the M67 CMD can be fitted with isochrones of models without overshoot if diffusion is included. Additionally, from spectroscopic results {\"O}nehag et al. (2014) claims M67 must have diffusion. Isochrones with varying degrees of diffusion and overshoot will be created for M67.

This article is organized as follows. Section 2 discusses the creation of the stellar models and isochrones, Section 3 describes the isochrone fitting to the CMD as well as the results, and Section 4 summarizes the best-fit parameters and provides concluding remarks.


\begin{table*}[t]
\centering
\caption{The model physics and the best fit values for the isochrones that were reasonable fits to the M67 CMD. These are the isochrone parameter combinations which were able to reproduce the M67 main sequence, turn-off, giant branch, and ``hook'' feature.}
\label{my-label}
\begin{tabular}{|c|c|c|c|c|c|c|c|c|}
\hline
\multicolumn{6}{|c|}{\textbf{Model Physics}} & \multicolumn{3}{c|}{\textbf{Best Fit Values}} \\ \hline
\multicolumn{1}{|c|}{\textbf{\begin{tabular}[c]{@{}c@{}}Row Reference\\ Number\end{tabular}}} & \multicolumn{1}{|c|}{\textbf{Diffusion}} & \multicolumn{1}{c|}{\textbf{Overshoot}} & \multicolumn{1}{c|}{\textbf{{[}Fe/H{]}}} & \multicolumn{1}{c|}{\textbf{$\alpha$}} & \multicolumn{1}{c|}{\textbf{$Y_0$}} & \multicolumn{1}{c|}{\textbf{\begin{tabular}[c]{@{}c@{}}Age\\ (Gyr)\end{tabular}}} & \multicolumn{1}{c|}{\textbf{\begin{tabular}[c]{@{}c@{}}Reddening\\ $E(B-V)$\end{tabular}}} & \multicolumn{1}{c|}{\textbf{\begin{tabular}[c]{@{}c@{}}Distance\\ Modulus\end{tabular}}} \\ \hline
1 & Yes & None & \phantom{$-$}0.1 & 2.0266 & 0.267 & 4.8 & 0.000 & 9.616 \\ \hline
2 & Yes & None & \phantom{$-$}0.1 & 2.0266 & 0.320 & 4.2 & 0.022 & 9.495 \\ \hline
3 & Yes & None & $-$0.1 & 2.0266 & 0.320 & 3.7 & 0.096 & 9.398 \\ \hline
4 & Yes & 0.2 & \phantom{$-$}0.1 & 2.0266 & 0.248 & 3.9 & 0.015 & 9.693 \\ \hline
5 & Yes & 0.2 & \phantom{$-$}0.0 & 2.0266 & 0.320 & 3.6 & 0.062 & 9.527 \\ \hline
6 & Yes & 0.2 & $-$0.1 & 2.0266 & 0.320 & 3.8 & 0.087 & 9.390 \\ \hline
7 & Yes & Mass-Based & \phantom{$-$}0.0 & 2.0266 & 0.248 & 4.4 & 0.035 & 9.608 \\ \hline
8 & Yes & Mass-Based & \phantom{$-$}0.0 & 2.0266 & 0.320 & 3.9 & 0.059 & 9.443 \\ \hline
9 & Yes & Mass-Based & $-$0.1 & 2.0266 & 0.248 & 4.1 & 0.071 & 9.520 \\ \hline
10 & Yes & Mass-Based & $-$0.1 & 2.0266 & 0.267 & 3.9 & 0.085 & 9.513 \\ \hline
\end{tabular}
\end{table*}

\section{Stellar Models}
\label{sec-1}
The stellar models were all created using the Yale Rotating Evolutionary Code (YREC) (Demarque et al., 2008). The OPAL equation of state (Rogers \& Nayfonov 2002) was used along with the OPAL high temperature opacities (Iglesias \& Rogers 1996) and low temperature opacities (Ferguson et al., 2005). The Adelberger et al. (1998) nuclear reaction rates were used, with the Formicola et al. (2004) value for the $^{14}\mathrm{N}(p,\gamma)^{15} \mathrm{O}$ reaction. In all cases the solar value of $Z_\odot/X_\odot=0.023$ was used (Grevesse \& Sauval 1998) and models were created using the Eddington T-$\tau$ relation in the atmosphere.

The models created spanned a wide range of parameter space. Models ranged in mass from $\mathrm{M}=0.65$ to 1.50 M$_\odot$, metallicity of $\mathrm{[Fe/H]}=-0.1$, 0.0, and 0.1, initial He abundance from $Y_{0}=0.248$ to 0.320, convective overshoot of 0.0, 0.2, and mass-dependent (from Demarque et al., 2004), and mixing lengths of $\alpha=1.5$ to 2.0. Additionally, models were created both with and without diffusion. The mass based overshoot, from Demarque et al. (2004), gives the following overshoot coefficient ($\Lambda_{\mathrm{os}}$) values: \\
$\Lambda_{\mathrm{os}} = 0.00$ for $M < M_\mathrm{crit}$, \\
$\Lambda_{\mathrm{os}} = 0.05$ for $M = M_\mathrm{crit}$, \\
$\Lambda_{\mathrm{os}} = 0.10$ for $M = M_\mathrm{crit} + 0.1$, \\
$\Lambda_{\mathrm{os}} = 0.15$ for $M = M_\mathrm{crit} + 0.2$, \\
$\Lambda_{\mathrm{os}} = 0.20$ for $M > M_\mathrm{crit} + 0.2$, \\
where $M$ and $M_{\mathrm{crit}}$ are in solar units. The value of $M_{\mathrm{crit}}$ depends on metallicity, as well as the composition of the star in terms of the $X$, $Y$, and $Z$ abundance ratios. For this set of models $M_{\mathrm{crit}}=1.2$ $\mathrm{M_{\odot}}$.

For the models with diffusion, the diffusion coefficient was set as 1 for masses below 1.25 M$_\odot$ and for masses larger than 1.25 M$_{\odot}$ was given by
\begin{equation}
\mathrm{diff \: coefficient}= \exp \left(\frac{- (M-1.25)^2}{2(0.085)^2}\right)
\end{equation}
In this way the value of the diffusion coefficient smoothly decreased for stars more massive than 1.25 M$_{\odot}$.


\subsection{Results}
\label{sec-2}
\subsection{Fitting Isochrones to the Cluster CMD}
For each set of parameters, isochrones for M67 were created at different ages. The best fit age, distance modulus, and reddening value for each set of parameters was then determined using 2-dimensional $\chi^2$ fitting between the points of the M67 CMD (Geller et al., 2015) and the isochrone. Only M67 stars with radial velocity membership probability of 90\% or greater were used (from Geller et al., 2015). Allowing the distance modulus, reddening, and age to be free parameters, a $\chi^2$ value for both the magnitude, $V$, and the color, $B-V$, was determined for each star in M67. These $\chi_{V}^{2}$ and $\chi_{(B-V)}^{2}$ values were summed to create the parameter $\chi_{\mathrm{total}}^{2}$. The age, reddening, and distance modulus combination that produced the lowest $\chi_{\mathrm{total}}^2$ was determined to be the best-fit isochrone for that model parameter set.

For each combination of model parameters, the best-fit isochrone was then visually inspected to determine if the isochrone was indeed a reasonable or realistic fit to the M67 cluster. In other words, that the isochrone fit the main sequence, turn off, giant branch, and reproduce the ``hook'' in the CMD. There were 10 best-fit isochrones that were reasonable fits to the M67 CMD. The properties of these isochrones are listed in Table 1 and they are plotted along with the M67 CMD in Figure 2.

\subsection{Effects of Model Physics on the M67 ``Hook''}
From the generated isochrones we can also investigate the parameters which give rise to the hook-like feature in the M67 CMD. As expected, isochrones without any diffusion and without any convective overshoot do not reproduce the hook. Models with convective overshoot do reproduce the hook in the CMD. However, we see that overshoot is not required to reproduce the hook if diffusion is included in the models. This confirms Michaud et al. (2014) claim that convective overshoot is not needed to model M67 if diffusion is implemented. A comparison of different isochrone properties and their effects on creating a hook-feature can be seen in Figure 3. In Figure 3 the isochrones are identical except with respect to their convective overshoot and diffusion amounts. The blue isochrone has neither overshoot or diffusion and fails to reproduce the hook feature. The green isochrone has no diffusion but does include convective overshoot and as a result has a hook in the CMD. The red isochrone does not have convective overshoot but does implement diffusion and also is able to reproduce this hook-like feature.

\begin{figure*}[ht]
\centering
\includegraphics[width=0.8\textwidth,clip]{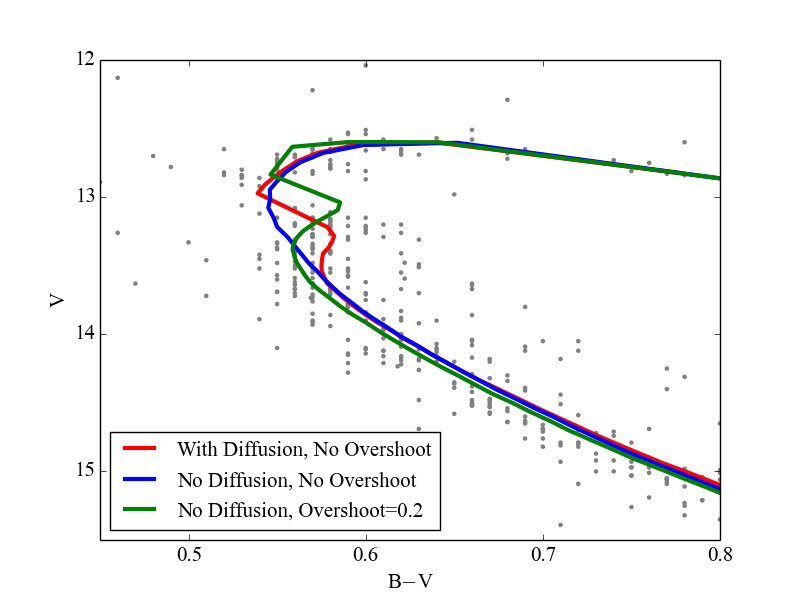}
\caption{A closer look at reproducing the M67 hook-like CMD feature. The red isochrone includes diffusion but no overshoot, the blue isochrone has no diffusion and no overshoot, and the green isochrone has no diffusion but does have convective overshoot. All other parameters of the isochrones are identical. As can be seen, both the presence of overshoot or the presence of diffusion can reproduce this hook-like feature while an isochrone with no diffusion and no overshoot cannot.}
\label{fig-3}       
\end{figure*}

\section{Conclusions}
\label{sec-con}

From the isochrones that fit the cluster, as seen in Table 1, we can get an age, distance, and reddening range. This gives an age of between 3.6 and 4.8 Gyr, a distance between 755 and 868 pc, and a reddening between $E(B-V)=0.000$ and 0.096. These ranges all agree well with the M67 parameter values from literature discussed in Section 1.

The best-fit isochrone has an age of 4.4 Gyr, distance of 834.9 pc, metallicity of $\mathrm{[Fe/H]}=0.0$, and reddening of $E(B-V)=0.035$. The physics of the best fit isochrone is with diffusion included, a mass-based overshoot, a mixing length of $\alpha=2.0266$, and an initial He abundance of $Y_{0}=0.248$.

We have also demonstrated the importance of allowing parameters that are generally held constant to vary and the effect that convective overshoot and diffusion can have on isochrone behavior. These results provide a good basis for parameter estimates which can aid in future asteroseismic studies of the cluster, for example helping to estimate expected frequencies of target stars. 

\section{Acknowledgments}
This work has been supported by NSF grant AST-1514676 and NASA grant NNX16AI09G to SB.

\begin{thebibliography}{19}

\bibitem{Adelberger1998}
{Adelberger}, E. G., {Austin}, S. M., {Bahcall}, J. N., et~al., Reviews of Modern Physics \textbf{70}, 1265 (1998)

\bibitem{Bohm1958}
{B{\"o}hm-Vitense}, E., ZAp \textbf{46}, 108 (1958)

\bibitem{PARSEC2012}
{Bressan}, A., {Marigo}, P., {Girardi}, L., et~al., MNRAS \textbf{427}, 127 (2012)

\bibitem{Demarque2008}
{Demarque}, P., {Guenther}, D. B., {Li}, L. H., {Mazumdar}, A., \& {Straka}, C.W., Ap\&SS \textbf{316}, 31 (2008)
 
\bibitem{Demarque2004}
{Demarque}, P., {Woo}, J.-H., {Kim}, Y.-C., \& {Yi}, S. K., ApJS \textbf{155}, 667 (2004) 
 
\bibitem{Ferguson2005}
{Ferguson}, J. W., {Alexander}, D. R., {Allard}, F., et~al., ApJ \textbf{623}, 585 (2005) 
 
\bibitem{Formicola2004}
{Formicola}, A., {Imbriani}, G., {Costantini}, H., et~al., Physics Letters B \textbf{591}, 61 (2004) 
  
\bibitem{Geller2015}
{Geller}, A. M., {Latham}, D. W., \& {Mathieu}, R. D., AJ \textbf{150}, 97 (2015)

\bibitem{Grevesse1998}
{Grevesse}, N. \& {Sauval}, A. J., Space Sci. Rev. \textbf{85}, 161 (1998)

\bibitem{Heiter2014}
{Heiter}, U., {Soubiran}, C., {Netopil}, M., \& {Paunzen}, E., A\&A \textbf{561}, A93 (2014)

\bibitem{Iglesias1996}
{Iglesias}, C. A. \& {Rogers}, F. J., ApJ \textbf{464}, 943 (1996)

\bibitem{Michaud2004}
{Michaud}, G., {Richard}, O., {Richer}, J., \& {VandenBerg}, D. A., ApJ \textbf{606}, 452 (2004)

\bibitem{Onehag2014}
{{\"O}nehag}, A., {Gustafsson}, B., \& {Korn}, A., A\&A \textbf{562}, A102 (2014)

\bibitem{Basti2004}
{Pietrinferni}, A., {Cassisi}, S., {Salaris}, M., \& {Castelli}, F., ApJ \textbf{612}, 168 (2004)

\bibitem{Richer1998}
{Richer}, H. B., {Fahlman}, G. G., {Rosvick}, J., \& {Ibata}, R., ApJL \textbf{504}, L91 (1998)
  
\bibitem{Rogers2002}
{Rogers}, F. J. \& {Nayfonov}, A., ApJ \textbf{576}, 1064 (2002)  

\bibitem{Sarajendini2009}
{Sarajedini}, A., {Dotter}, A., \& {Kirkpatrick}, A., ApJ \textbf{698}, 1872 (2009)

\bibitem{Taylor2007}
{Taylor}, B. J. AJ \textbf{133}, 370 (2007)

\bibitem{Yi2001}
{Yi}, S., {Demarque}, P., {Kim}, Y.-C., et~al., ApJS \textbf{136}, 417 (2001)

\end{thebibliography}

\end{document}